\documentclass[aps,prl,preprint,showpacs,groupedaddress,floatfix]{revtex4}
\usepackage{graphicx}
\usepackage{amssymb,amsmath,braket}
\usepackage{ulem}
\usepackage[dvips]{color}

\begin{document}

\title{Poisson's Ratio of Layered Two-dimensional Crystals}
\author{Sungjong Woo}
\author{Hee Chul Park}
\author{Young-Woo Son}
\affiliation{Korea Institute for Advanced Study, Seoul 130-722, Korea}

\date{\today}	
\begin{abstract}
We present first-principles calculations of elastic properties of multilayered 
two-dimensional crystals such as graphene, $h$-BN and $2H$-MoS$_2$
which shows that their Poisson's ratios along out-of-plane direction
are negative, near zero and positive, respectively, 
spanning all possibilities for sign of the ratios.
While the in-plane Poisson's ratios are all positive regardless of their disparate electronic 
and structural properties, 
the characteristic interlayer interactions as well as layer stacking structures 
are shown to determine the sign of their out-of-plane ratios.
Thorough investigation of elastic properties as a function of the number of layers
for each system is also provided, highlighting their intertwined nature
between elastic and electronic properties.
\end{abstract}

\pacs{62.20.de, 62.20.dj, 73.21.Ac}

\maketitle

Under uniaxial stress, Poisson's ratio defined by the ratio 
of the strain in the transverse direction ($\epsilon_{t}$)
to that of the longitudinal direction ($\epsilon_{l}$), $\nu=-\epsilon_{t}/\epsilon_{l}$,
measures the fundamental mechanical responses of solids against 
external loads~\cite{Landau,Rouxel,Poole,Greaves2,Greaves}. 
It has strong correlation with atomic packing density, atomic connectivity~\cite{Rouxel} and 
structural phase transition~\cite{Poole, Greaves2, Greaves}.
The theory of elasticity allows values of Poisson's ratio of an isotropic material ranging 
from $-1$ to 0.5, i.e., from extremely compressible to incompressible
materials~\cite{Landau,Greaves}.
Thus, when a solid is subjected to a uniaxial compression, 
it expands ($\nu>0$), remains to be the same ($\nu=0$), and shrinks ($\nu<0$) 
in the transverse direction depending on the sign of Poisson's ratio.
Typically, different Poisson's ratio or its sign indicates dramatic 
variations in mechanical properties. 
For example, when isothermal modulus is extremely larger than shear modulus, 
the material reaches its incompressible limit as shown in most liquids or rubber ($\nu\sim0.5$)
and in the opposite case, re-entrant foams and related structures show the negative $\nu$
or auxetic property~\cite{Greaves,Lakes, Caddock, Milton,Evans}.
The Poisson's ratio of common solid state crystals usually falls in the range of $0<\nu <0.5$ while
gases and cork have $\nu \simeq 0$~\cite{Greaves,Lakes, Caddock, Milton,Evans}.

Anisotropic materials with directional elastic properties often shows more 
dramatic variations in their Poisson's ratios such as 
the directional auxetic property~\cite{Greaves}.
In this regard, the experimental realization of graphene~\cite{NovoselovGeim, ZhangKim}, 
the thinnest and the strongest material~\cite{Lee, Marianetti,si,sjwoo}, 
now offers a new platform to understand electronic and elastic properties
of well-defined anisotropic materials and their heterostructures.
Even though the Young's modulus
and Poisson's ratio of graphene have been studied quite thoroughly
\cite{Lee, Marianetti,si,sjwoo,Koenig, Zhang, Kordkheili, Blackslee, Politano, Scarpa}, 
those along the out-of-plane 
direction for its few-layered forms have barely been known.
Neither do for all the other available two-dimensional crystals.
Since electronic properties of layered two-dimensional crystals vary a lot depending
on their chemical composition as well as the number of layers~\cite{NovoselovGeim2, GeimNovoselov,Geim}, 
their corresponding elastic properties, especially for few layered structures, are
anticipated to change accordingly. 
Motivated by recent rapid progress in manipulating various two-dimensional crystals
and their stacking structures~\cite{GeimNovoselov,Geim,CRDean}, we have calculated
fundamental mechanical properties of three representative
van der Waals (vdW) crystals along all crystallographic directions of their few-layered structures. 

In this work, we present a theoretical study using a first-principles approach 
on the elastic properties of layered two-dimensional crystals, 
including graphene, $h$-BN and 2$H$-MoS$_2$,
in which the vdW energy is one of the governing interactions between their layers
while they exhibit very different electronic properties.
We find that the Poisson's ratios of graphene, $h$-BN and 2$H$-MoS$_2$ along out-of-plane 
direction are negative, near zero and positive, respectively, whereas their in-plane Poisson's ratios
are all positive.
The diverseness of out-of-plane Poisson's ratio is attributed to their disparate electronic
properties as well as stacking structures.
Thorough investigation on their elastic properties  
while varying the number of layers are also reported.

We first consider graphene with $AB$ stacking, 
$h$-BN and 2$H$-MoS$_2$ with $AA'$ stacking. 
All the three have $C_{3v}$ symmetry.
Generally, for a material with $C_{3v}$ symmetry, the stiffness tensor without shear part
can be written with four independent parameters,
\begin{equation}
\label{stiffness}
\begin{pmatrix} \sigma_{x} \\ \sigma_{y} \\ \sigma_{z} \end{pmatrix} =  
\begin{pmatrix} A & B & C \\ B & A & C \\ C & C & D \end{pmatrix}
\begin{pmatrix} \epsilon_{x} \\ \epsilon_{y} \\ \epsilon_{z} \end{pmatrix},
\end{equation}
with the choice of $z$ as the axis for the three-fold rotational symmetry~\cite{Landau}.
Here, $\sigma_i$ and $\epsilon_i$ are the stress and strain respectively 
along the $i$-th axis.
The components of stiffness tensor can be obtained by differentiating the total energy
$E_{\rm tot}$ in terms of strain; 
$A=\partial^2 E_{\rm tot}/\partial \epsilon_{x}^2
=\partial^2 E_{\rm tot}/\partial \epsilon_{y}^2$, 
$B=\partial^2 E_{\rm tot}/\partial \epsilon_{x} \partial \epsilon_{y}$, 
$C=\partial^2 E_{\rm tot}/\partial \epsilon_{x} \partial \epsilon_{z}$, 
and $D=\partial^2 E_{\rm tot}/\partial \epsilon_{z}^2$.
By taking the inverse of the stiffness tensor, one can get the compliance
tensor,
\begin{equation}
\label{compliance}
\begin{pmatrix} \epsilon_{x} \\ \epsilon_{y} \\ \epsilon_{z} \end{pmatrix} = 
\begin{pmatrix} 1/E_i & -\nu_i/E_i & -\tilde{\nu}_o/E_o \\ -\nu_i/E_i & 1/E_i & -\tilde{\nu}_o/E_o \\ -\nu_o/E_i & -\nu_o/E_i & 1/E_o \end{pmatrix}
\begin{pmatrix} \sigma_{x} \\ \sigma_{y} \\ \sigma_{z} \end{pmatrix}.
\end{equation}
The subscripts $i$ and $o$ represent {\it in-plane} and {\it out-of-plane} respectively.
$E_i$ and $E_o$ are the Young's moduli along the $x(y)$ and $z$ axis respectively.
There are two out-of-plane Poisson's ratios; $\nu_o$ is the Poisson's ratio along the $z$ axis when the stress is applied
along the $x$ or $y$ directions while $\tilde{\nu}_o=\nu_0 E_0/E_i$ is the Poisson's ratio
along the $x$ or $y$ direction when the stress is applied along the $z$ direction.
$\nu_i$ is the in-plane Poisson's ratio along the $x (y)$ axis when the stress is applied 
along the $y (x)$ axis.

\begin{figure}
	 \centerline{\includegraphics[width=1\columnwidth]{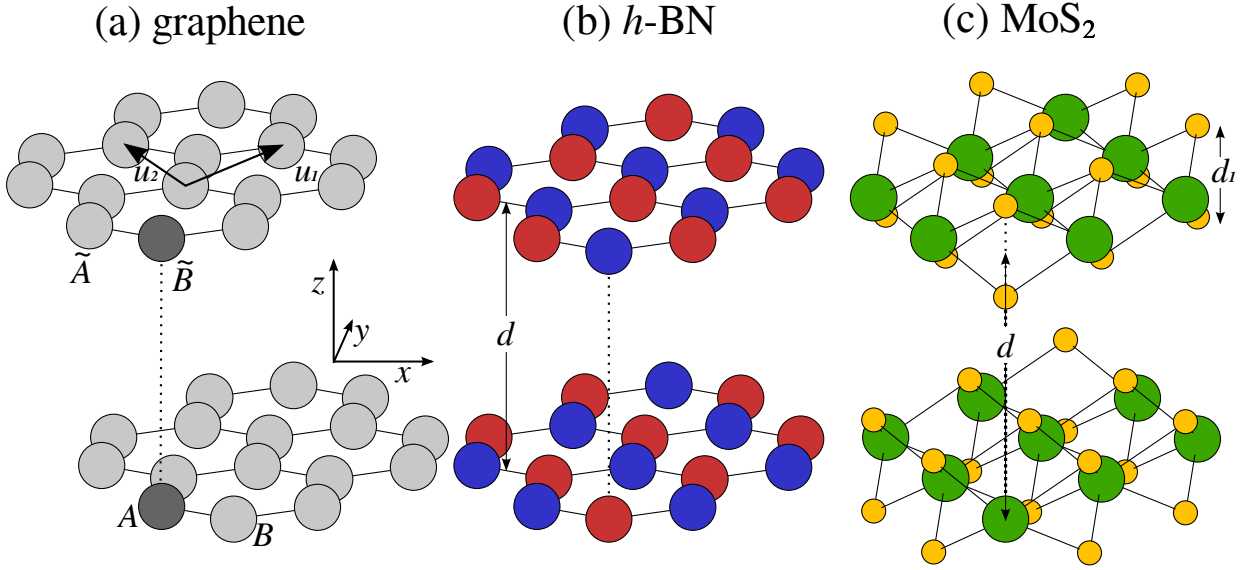}}
 \caption{
 Lattice structures of (a)~$AB$-stacked 
 graphene (b)~$AA'$-stacked $h$-BN and (c)~$2H$-MoS$_2$. 
 The parameter $d$ 
 is the interlayer distance of each structure.
 In $2H$-MoS$_2$, $d$ is the vertical distance between Mo atoms in adjacent layers
 and  $d_1$ is the vertical intralayer sulfur-to-sulfur distance.}
 \label{Fig1}
\end{figure}

Using a first-principles approach
based on density-functional theory with plane wave basis set~\cite{QE},
we calculate total energies, $E_{\rm tot}(\epsilon_{x}, \epsilon_{y}, \epsilon_{z})$,
of all systems at $5\times 5\times 5$ 
grid points in the strain space of $(\epsilon_{x}, \epsilon_{y}, \epsilon_{z})$.
To obtain the accurate binding energy and interlayer distance including the vdW energy, 
we have used the revised version~\cite{Sabatini} 
of the nonlocal correlation functional method developed
by Vydrov and van Voorhis~\cite{VV10}
that is successful for reproducing both values following 
results from more accurate methods~\cite{Bjorkman}.
In order to reduce spurious interaction between neighbouring supercells,
a large vacuum over 68~\AA~is introduced
and relatively high energy cut-off above 100 Rydberg 
as well as dense k-point grids up to $29\times 29$ are used to converge the results. 

For the total energy calculations with tensile strain on all systems
in which the layers are stacked along the $z$ axis,
a primitive cell with unit vectors, 
$u_1=(a, b, 0)$, $u_2=(-a, b, 0)$, $u_3=(0, 0, c)$ are used~[Fig.~\ref{Fig1}].
Strain along $x$ and $y$ axis is defined by 
$\epsilon_{x} = (a-a_0)/a_0$
and $\epsilon_{y} = (b-b_0)/b_0$,
where $a_0$ and $b_0$ are the lattice parameters of the equilibrium structure.
Strain across the layers along the $z$ axis 
is defined by $\epsilon_{z} = (d-d_0)/d_0$ 
where $d$ and $d_0$ are
the interlayer distance and that of the equilibrium structure, respectively.
The calculated lattice parameters of $a_0$ and $d_0$ for infinitely stacked bulk
systems are 2.47 \AA, 2.52 \AA, 3.22 \AA~and 6.72 \AA, 6.61 \AA, 12.42 \AA~for graphite, $h$-BN, 
and $2H$-MoS$_2$, respectively, which are in excellent agreements with
previous studies~\citep{Baskin, Paszkowicz, Dickinson, Jellinek}.
The slight variation of $a_0$, $b_0$ and $d_0$ depending on the number of layers 
are reflected in our calculations.
$A$, $B$, $C$ and $D$ in Eq. (1) are calculated by interpolating $E_{\rm tot}$ 
on the strain space and
Young's modulus and Poisson's ratio from compliance tensor in Eq. (2). 

\begin{figure}
 \centerline{\includegraphics[width=1\columnwidth]{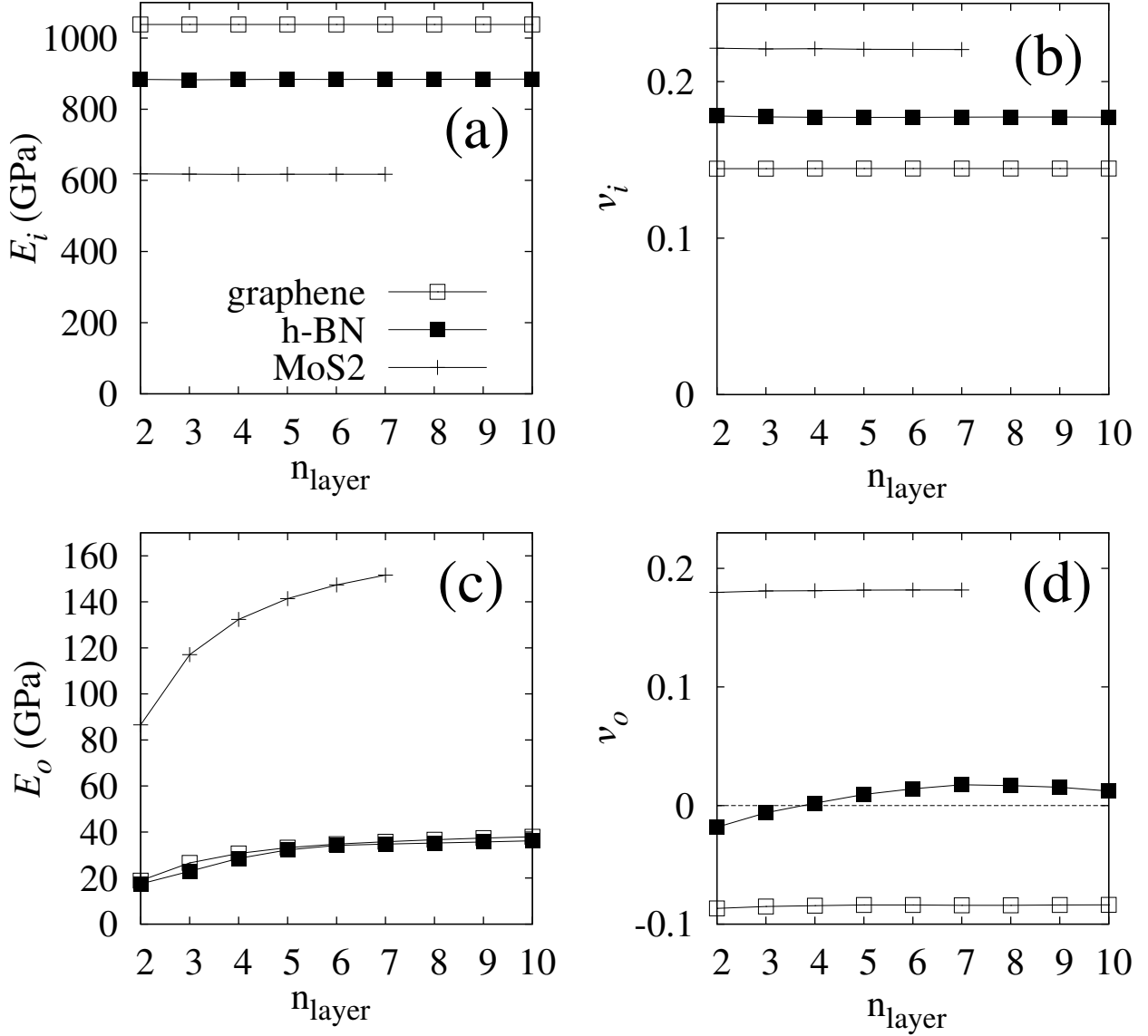}}
 \caption{
 Elasticity constants of graphene (empty 
 rectangle), $h$-BN (filled rectangle) and 
 $2H$-MoS$_2$ (cross) as a function of number of layers upto ten 
 layers, in-plane (a) Young's modulus and (b) Poisson's ratio and 
 out-of-plane (c) Young's modulus and (d) Poisson's ratio.}
 \label{Fig2}
\end{figure}

Figure \ref{Fig2} shows Young's moduli and Poisson's ratios for the three
materials with various number of layers.
In-plane elastic constants, $E_i$ and $\nu_i$, are barely dependent on 
the number of layers~[Figs.~\ref{Fig2}(a) and (b)].
Under in-plane tensile stress, the hexagonal network of atoms is
deformed for graphene and $h$-BN while, for $2H$-MoS$_2$, sulfur-to-sulfur distance 
across the plane along the $z$ axis within one layer can also be deformed.
Furthermore, the hexagonal structure of graphene and $h$-BN with rigid 
$\sigma$ bond supported by $\pi$ bond is stiffer than that of $2H$-MoS$_2$.
This makes $2H$-MoS$_2$ more flexible to applied stress 
resulting in lower in-plane Young's modulus~[Fig.~\ref{Fig2}(a)] and 
larger in-plane Poisson's ratio~[Fig.~\ref{Fig2}(b)].
Young's moduli across layers, $E_o$, increase for all of the three materials with
the increase of the number of layers reflecting the accumulation of long-range interlayer
van der Waals interaction~[Fig.~\ref{Fig2}(c)].

Contrary to similar behaviours between in-plane elastic properties of the three materials, 
out-of-plane elastic properties between those differ qualitatively [Fig. 2(d)]. 
The most notable one is 
that multilayered graphene structures have out-of-plane Poisson's ratio 
as negative as $\nu=-0.09$~[Fig.~\ref{Fig2}~(d)] with slight dependence on 
the layer number variations.
Materials with axial negative Poisson's ratio 
have been reported during the last few decades such as
foams
with re-entrant atomic structures~\cite{Greaves,Lakes, Caddock, Milton, Evans}
and those with non-axial one are shown in 
some simple cubic metals~\cite{Huang, Baughman}.
The present case is for the axial negative ratio in a layered material where
the vdW interaction governs the binding between layers
without re-entrant structure.
More interestingly, $h$-BN shows very small out-of-plane Poisson's ratio near zero crossing
from negative to positive values as number of layers increases whereas 2$H$-MoS$_2$ 
has positive Poisson's ratio as shown in Fig. 2(d).
So, the three layered crystals have qualitatively different Poisson's ratios spanning all 
possibilities of their signs.

\begin{figure}
 \centerline{\includegraphics[width=1\columnwidth]{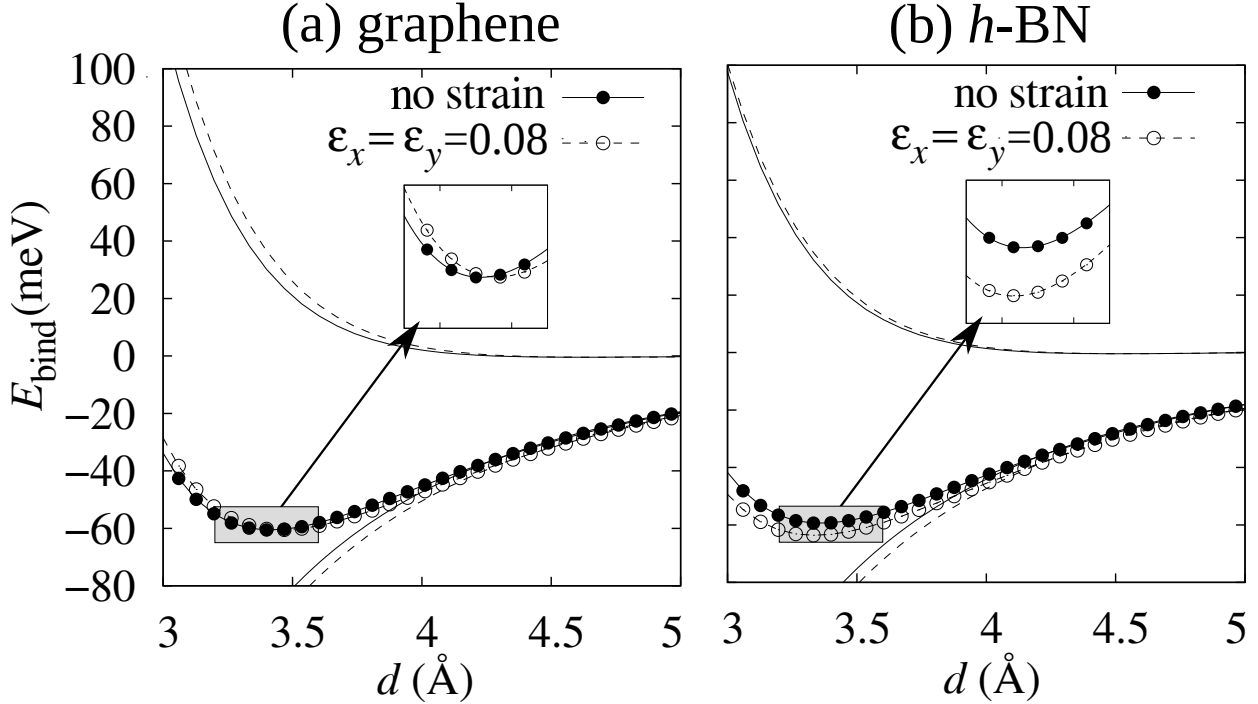}}
 \caption{
Binding 
energy as a function of interlayer distance, $d$, for bilayer 
 (a) graphene and (b) $h$-BN. 
 The fitting curves for attractive and repulsive parts are drawn in for unstrained case (solid line) 
 and $\epsilon_x=\epsilon_y=0.08$ equibiaxial strain case (dashed 
 line).
 The insets show magnified views near equilibrium.
 }
 \label{Fig3}
\end{figure}

To understand the qualitative difference in out-of-plane Poisson's ratios of the three layered systems, 
we first decompose the binding energy of bilayer systems into repulsive and attractive parts. 
Figure \ref{Fig3}(a) shows
the binding energy curve between two layers of graphene, 
$E_{\rm bind}(d) \equiv \frac{1}{2}\left[E_{\rm tot}^{\rm b}(d)-2E_{\rm tot}^{\rm s}\right]$
where $E_{\rm tot}^{\rm b(s)}$ is the total energy of bilayer (single layer) graphene.
$E_{\rm bind}(d)$ is shown as solid (no strain) and
open ($8.1\%$ equibiaxial nominal strain) circles as a function of the 
distance, $d$, between the layers. 
The calculated binding energy within $d=4\sim 9~{\rm\AA}$ 
is well described by $E_{\rm bind}(d)\sim d^{-4}$.
The fitting curves of $d^{-4}$, which reflect asymptotic vdW interaction, 
are drawn in as a solid (dashed) line without (with) strain.
The difference between the total energy and vdW energy for each case 
is also plotted in the same plot, two curves on top, 
representing purely repulsive characteristics called  
Pauli repulsion~\citep{Jensen, Gritsenko}.
It does not fit to any single power of $d^{-\alpha}$ 
but is well fit by 
an exponentially decaying function,
$E_{\rm bind}(d) \sim \left[\exp(d^2/\sigma^2)-1\right]^{-1}$,
with $\sigma=1.37~{\rm \AA}$ 
for the case without strain.
We note that the $d^{-3}$ dependence of vdW energy of 
bilayer graphene,
which was recently reported~\cite{Dobson} is valid only at a distance 
larger than $9~{\rm \AA}$~\cite{Lebegue} therefore not relevant 
near equilibrium distance considered here.

Figure \ref{Fig3}(a) indicates that both vdW attraction and Pauli repulsion
on bilayer graphene are enhanced under tensile strain.
However, noting that the Pauli repulsion energy showing exponential increase with $d$ is much stiffer
than attractive vdW energy,
equilibrium interlayer distance is critically sensitive to
the change of the former than the latter.
Thus, the equilibrium interlayer distance under the strain
is mainly determined not by the strain-enhanced vdW attraction~\cite{Sharma}
but by the enhancement of the repulsion.
In graphene, electronic states pointing away from the layers 
are composed of linear combinations of $p_z$ orbitals of atoms called $\pi$ orbitals and
form the $\pi$ band~\cite{Saito}.
Since the occupied electrons of $\pi$ orbitals in adjacent layers expel each other from their overlap region~\cite{Chakraborty}, 
the enhanced repulsion with external strain shown in Fig. \ref{Fig3} (a) may indicate 
the strain-induced spatial variation of $\pi$ electrons pushing the two layers away while the vdW interaction
still keeps their binding.

We find that the in-plane strain indeed elongates the spatial distribution of electron density 
away from the layer making the Pauli repulsion increase 
over the vdW attraction.
We calculate the spatial distribution of density of $\pi$ band
along the $z$ axis in a single layer graphene;
$
  \rho_\pi(z) \equiv \int^{\varepsilon_F}_{-\infty} dE \int dx dy\rho_E(x,y,z) .
$
Here, $\rho_E(x,y,z)$ is the local density of state for the $\pi$ band only and 
is summed over the in-plane unit cell 
($x$ and $y$ axis).
Graphene is located at $z=0$ and the Fermi energy of the neutral system is denoted 
by $\varepsilon_F$.
Our {\it ab initio} calculation result for the maximum of $\rho_\pi(z)$
decreases with tensile strain while 
its tail 
increases implying that $\pi$ orbital spreads out along the $z$ axis with strain.
Quantitatively, we calculate the density-weighted length of $\pi$ orbital along the $z$ axis using
$L_{\pi} \equiv \int~dz~|z|\rho_\pi (z) / 
\int~dz~\rho_\pi (z)$ that gives $L_\pi = 0.673~{\rm \AA}$ without strain.
We find that the value of $L_{\pi}$ indeed increases by 0.6 \% as the equibiaxial strain increases by 2 \%, 
thereby explaining the value of the negative Poisson's ratio near $-0.1$ along out-of-plane direction.
This elongation can be understood simply by considering overlaps between 
neighbourig atomic orbitals. 
For a charge neutral graphene, 
the spatial distribution of $\pi$ orbitals along the perpendicular direction to the layer 
is contracted compared to $p_z$ orbitals of an isolated carbon atom because of overlap between
nearby $p_z$ orbitals.
In-plain tensile strain returns the carbon atoms in graphene back to isolated one so that the 
$\pi$ orbitals should be elongated.

A simple tight-binding (TB) picture can corroborate the elongation of spatial distribution of $\pi$ orbitals under strain.
Consider the Bloch wave function within the TB approximation,
$\phi_{A(B)}(\vec{k}, \vec{r})=N^{-1/2}\sum_{\vec{R}_{A(B)}}e^{i\vec{k}\cdot \vec{R}_{A(B)}}\varphi_{A(B)}(\vec{r}-\vec{R}_{A(B)})$,
where subscript $A(B)$ represents the sublattice index, 
$\varphi_{A(B)}(\vec{r}-\vec{R}_{A(B)})$ is the normalized $p_z$ orbital of the carbon atom at $\vec{R}_{A(B)}$,
while $\vec{R}_{A(B)}$ runs the positions of atoms in the $A(B)$ sublattice~\citep{Saito}.
With the nearest neighbor hopping, $t$,
and the overlap, $s=\braket{\varphi_A({\vec r}-{\vec R}_A)|\varphi_B({\vec r}-{\vec R}_A-{\vec\delta}_j)}$, 
the $\pi$-orbital is given by
$\psi_\pi(\vec{k},\vec{r}) 
= \frac{f(\vec{k})}{|f(\vec{k})|}\phi_A(\vec{k},\vec{r}) + \phi_B(\vec{k},\vec{r})$ 
where $f(\vec{k})=\sum_{j=1}^3 e^{i\vec{k}\cdot\vec{\delta}_j}$
and $\vec{\delta}_j$ points to the three nearest neighbors.
Considering $s\ll 1$, 
$L_{\pi} = \frac{1}{S_{\rm BZ}}\int_{\rm BZ} d^2k ~l_{\pi}(\vec{k}) $,
where
$l_{\pi}(\vec{k})\equiv\frac{\langle\psi_\pi (\vec{k})| |z| |\psi_\pi (\vec{k}) \rangle}{\langle\psi_\pi (\vec{k})|\psi_\pi (\vec{k}) \rangle}
\approx l_{p_z} - |f(\vec{k})|(l_{p_z}s - l_\delta)$ is 
the density-weighted length of the $\pi$ orbital at $\vec{k}$
and $S_{\rm BZ}$ is the area of the first Brillouin zone (BZ).
Here, $l_{p_z}=\int d^3r~|z||\varphi_{A}(\vec{r})|^2$ is the
length of an isolated $p_z$ orbital and
$l_\delta=\int d^3r~\varphi_A^*(\vec{r}-\vec{R}_A) |z|\varphi_B({\vec r}-\vec{R}_A-\vec{\delta}_1)$.
If the maximum overlap between nearest neighbor $p_z$ orbitals is
at $|z|=z_0$ and
$\varphi_A^*(\vec{r}-\vec{R}_A)\varphi_B(\vec{r}-\vec{R}_A-\vec{\delta}_1)$
is trivial elsewhere, 
then $l_\delta\approx z_0 s$ 
so that $l_\pi(\vec{k}) \approx l_{p_z} - |f(\vec{k})|(l_{p_z}-z_0)s $
and that $L_\pi \approx l_{p_z}-l_{h}(|\vec{\delta}_1|)$
where $l_{h}\equiv\frac{(l_{p_z}-z_0)s}{S_{\rm BZ}}\int_{\rm BZ} d^2k ~|f(\vec{k})| $.
It is straightforward to find that 
$l_h > 0$ and $\partial l_h / \partial |\vec{\delta}_1| <0$.
Therefore, the above simple formulation for $L_{\pi}$
implies that the out-of-plane distance of $\pi$-orbital is shorter than that of bare $p_z$ orbital
and that the applied tensile strain can increase its distance.

Now, let us compare elastic properties of multilayered $h$-BN with graphene.
A previous study \cite{Hod} shows that the ionic interaction energy in the $h$-BN
is negligible in determining the electrostatic repulsion as well as dispersion forces
so that the interlayer distance is very similar to graphite
regardless of apparent difference in the static polarizability between the two layered materials.
Our analysis, however, shows the elastic properties can be quite different; 
amplitude of out-of-plane Poisson's ratio of $h$-BN 
is nowhere close to that of multilayer graphene but order of magnitude
smaller [Fig.~\ref{Fig2}(d)].
Figure~\ref{Fig3}(b) shows that under tensile strain, vdW interaction
increases as in graphene while the Pauli repulsion barely changes.
For a bilayer graphene with $AB$ stacking, half of the carbon atoms
of one layer are right on top of the carbon atoms of the other layer
so that the tails of $p_z$ orbitals from two layers directly overlap with each other.
For a bilayer $h$-BN with $AA'$ stacking, however, fully-filled $p_z$
orbital of nitrogen atom from one layer is on top of the empty $p_z$ orbital
of boron from the other layer so that the interlayer Puali repulsion
is not as sensitive to the slight change of the length of $p_z$ orbitals
as in the case of graphene.

The length of $p_z$ orbital of a single-layer $h$-BN, in fact, does
change due to the strain in the same manner as that of graphene.
Therefore, a negative interlayer Poisson's ratio should appear in $h$-BN as
multilayer graphene if the interlayer alignment of $p_z$ orbitals 
follows that of graphene.
This can be realized by changing the stacking structure of $h$-BN from $AA'$ to $AA$.
We have computed the elastic properties for this artificial bilayer structure
and found that the out-of-plane 
Poisson ratio is $-0.12$, thus confirming our theory.
On the other hand, the out-of-plane Poisson's ratio of bilayer 2$H$-MoS$_2$ is 
mainly determined by the flattening of each layer under tensile strain that gives
positive value.
Our calculation shows that the change of interlayer distance, $d$, of 2$H$-MoS$_2$ 
in Fig.~\ref{Fig1}(c) 
in response to a given in-plane tensile stress mainly comes from the change of $d_1$;
$\Delta d_1 \approx (3/4)\Delta d$.

In conclusion, we have studied the elastic properties of multilayered two-dimensional
crystals including graphene, $h$-BN, and 2$H$-MoS$_2$, with interlayer van der Waals
interaction properly taken into account.
In-plane elastic properties are found to be barely dependent on the number 
of layers for all three materials.
Our analysis reveals that graphene is a very peculiar axial auxetic material when in-plain
strain is applied.
The mechanism is attributed to quantum mechanical origin rather than to structural
one such as re-entrant foam.
In contrast, the Poisson's ratio of $h$-BN with $AA'$ stacking is found to be nearly zero 
and that of MoS$_2$ is positive.

\begin{acknowledgements}
We thank Jae-Hyun Kim and Yun Hwangbo for fruitful discussions at early stage of this study.
Y.-W.S. was supported by the NRF funded by the MSIP 
of Korean government (CASE, 2011-0031640 and QMMRC, No. R11-2008-053-01002-0).
Computations were supported by the CAC of KIAS.
\end{acknowledgements}

\end{document}